\input{aipcheck}

\documentclass[
    ,final            
  ]
  {aipproc}

\layoutstyle{6x9}

\begin{document}

\title{Simple theoretical models for resonant cold atom interactions}


\author{Paul S. Julienne}{
  address={Atomic Physics Division, NIST, Gaithersburg, Maryland 20899-8423 USA}
}

\author{Bo Gao}{
  address={Department of Physics and Astronomy, University of Toledo, Toledo, Ohio 43606}
}

\begin{abstract}
Magnetically tunable scattering resonances have been used with great success for precise control of $s$-wave scattering lengths in ultracold atomic collisions.  We describe relatively simple yet quite powerful analytic treatments of such resonances  based on the analytic properties of the van der Waals long range potential.  This theory can be used to characterize  a number of properties of specific resonances that have been used successfully in various experiments with $^{87}$Rb, $^{85}$Rb, $^{40}$K, and $^{6}$Li.  Optical Feshbach resonances are also possible and may be practical with narrow intercombination line photoassociative transitions in species like Sr  and Yb.

{\bf This will be published in {\it Atomic Physics 20}, the Proceedings of the 20th international Conference on Atomic Physics, Innsbruck, Austria, July 2006.} 
 \end{abstract}

\maketitle


\section{Introduction}

Tunable control of scattering resonances between interacting ultracold atoms has been widely used in many forefront experiments in atomic physics, many of which are described at this Conference.  Such experiments may use bosonic or fermionic species and explore phenomena such as Bose-Einstein condensation (BEC), quantum degenerate Fermi gases, cold molecule formation, molecular BEC, few-body physics, or reduced dimensional structures in optical lattices.  The control of such systems comes from the ability to tune the near-threshold scattering and bound state properties of interacting cold atoms.  Threshold collisions are characterized by the scattering length $a$, which represents a quantum phase expressed in units of length: the scattering phase shift $\eta(E)\to  -ka$ as $E\to 0$ at low collision energy $E$, where $\hbar k$ is the relative collision momentum.  The ability to make $a$ vary over a wide range between $\pm \infty$ using a tunable scattering resonance is what allows for precise control of cold atomic gases.  This talk will not survey the kinds of experiments that are being done, since these will be covered by many other speakers, but will instead concentrate on developing some simple  models that  provide understanding of these resonances, using as examples resonances that have been or are being exploited in key ultracold atom experiments.

In general, scattering resonances are complex and require either  detailed experimental studies in order to measure their properties or detailed knowledge of the molecular physics of the interacting system in order to calculate their properties.  On the other hand, they remain amenable to simple models.  The simplest is a {\em universal} model parameterized solely in terms of the scattering length $a$ and reduced mass $\mu$ of the pair of atoms.  This will apply to scattering and bound states sufficiently close to threshold.  However, the subject of this talk is a much more powerful model, extending over a much wider range of energy, based on the analytic properties of the van der Waals potential.  We will primarily consider resonances tunable by a magnetic field $B$, but briefly consider resonances tunable by an optical field of frequency $\nu$.

\section{Resonant scattering}

There is a long history of resonance scattering, going back to the earliest days of quantum physics~\cite{Rice33,Fano35}, and widely developed in nuclear physics, especially the elegant formalism of Herman Feshbach~\cite{Feshbach58}, whose name has become attached to the resonances used in ultracold atom research.  Tiesinga {\it et al.}~\cite{Tiesinga93} pointed out the existence of sharp resonances versus $B$ in the scattering length of ultracold Cs atoms, and the quest for resonance control of ultracold collisions was underway.  A recent review~\cite{Kohler06} describes the near threshold properties of magnetically tunable Feshbach resonances in the context of cold molecule formation and atom paring.  

Whatever theory one uses, the colliding two-atom system is separated into an approximate bound state $|n\rangle$ with discrete energy $E_n$ and a scattering continuum $|E\rangle$ with a continuous spectrum of collision energies $E$ with some coupling $V$ between $| n \rangle$ and $|E\rangle$. The "bare" bound state $|n\rangle$ in the absence of coupling is the closed channel, or resonance, state, and the "bare" continuum $|E\rangle$ in the absence of coupling represents the open channel, or "background" scattering.  Following Fano's classic 1961 paper~\cite{Fano61}, the scattering phase shift $\eta(E)$ separates into background and resonance components: $\eta(E)=\eta_\mathrm{bg}+\eta_\mathrm{res}(E)$, where the weak energy dependence of $\eta_\mathrm{bg}$ over a narrow resonance can be ignored.  The resonance contribution varies rapidly by $\pi$ as $E$ varies from below to above resonance at shifted energy $E_0=E_n+\delta E_n$:
\begin{equation}
 \eta_\mathrm{res}(E)=-\tan^{-1}\frac{\frac{1}{2}\Gamma_n}{E-E_n-\delta E_n}\,,
\end{equation}
where $\Gamma_n=2\pi|\langle n|V|E_n\rangle|^2$ and $\delta E_n$ are the resonance width and shift respectively. 

This theory is modified for near-threshold resonances in that  $\eta_\mathrm{bg}(E)$, $\Gamma_n(E)$, and $\delta E_n(E)$ become strongly $E$-dependent, following quantum threshold law behavior.  For a magnetically tunable resonance $E_n = \delta \mu (B-B_n)$ crosses threshold ($E=0$) at $B=B_n$, and $\delta \mu$ gives the magnetic moment difference between the "bare" open channel atoms and the "bare" resonance state.  The phase near threshold is:
\begin{equation}
 \eta_\mathrm{res}(E)=-\tan^{-1}\frac{\frac{1}{2}\Gamma_n(E)}{E-\delta \mu(B-B_n)-\delta E_n(E)}\,,
\end{equation}
As $E\to0$ the threshold relations $\eta_\mathrm{bg}\to-ka_\mathrm{bg}$ and $\frac{1}{2}\Gamma_n(E) = (ka_\mathrm{bg}) \delta \mu\,\Delta_n$~\cite{Mies00} imply a tunable scattering length
\begin{equation}
 a(B) = a_\mathrm{bg}\left ( 1 - \frac{\Delta_n}{B-B_0} \right )\,,
\end{equation}
where $\Delta_n$ is the width of the resonance in magnetic field units, and $a(B)$ is singular at the shifted resonance "position" $B_0=B_n+\delta B_n$, where $\delta B_n = \delta E_n/\delta\mu$.  If $a(B)$ is positive and sufficiently large, there is a bound state with binding energy $E_b$   related to $a(B)$ by the "universal" relation $E_b=\hbar^2/(2 \mu a(B)^2)$.  This universal relation applies for a number of known experimental resonances~\cite{Kohler06}.

\section{Analytic van der Waals theory}

While universal resonance properties parameterized by $a(B)$ are certainly useful, a much more powerful theory is possible by introducing the analytic properties of the van der Waals potential, which varies at large interatomic separation $R$ as $-C_6/R^6$.  The solutions for a single potential, depending solely on $C_6$, reduced mass $\mu$ and scattering length $a$ for that potential, have been worked out in a series of articles by B. Gao~\cite{Gao98a,Gao98b,Gao99,Gao00,Gao01,Gao04a}.   The van der Waals theory is especially powerful for threshold resonances and bound states when the van der Waals solutions are used as the reference solutions for the specific form of the multichannel quantum defect theory (MQDT) developed by Mies and coworkers~\cite{Mies84a,Mies84b,Julienne89,Mies00a}.  In particular, the MQDT is concerned with the analytic properties of single and coupled channels wave functions $\Psi(R,E)$ across thresholds as $E$ goes from positive to negative and between short and long range in $R$.  The van der Waals and MQDT theories, when combined, give simple formulas for threshold resonance properties that have illuminating physical interpretations.   For all cases where we have tested it numerically, the van der Waals MQDT gives scattering properties in excellent agreement with full coupled channels scattering calculations over a wide range of energies exceeding those normally encountered in cold atom experiments below 1 mK.  It is especially good in the ultracold domain of a few $\mu$K and below.

The key parameters are the scale length $R_\mathrm{vdw}=(1/2)\left (2\mu C_6/\hbar^2 \right)^{1/4}$ and corresponding energy $E_\mathrm{vdw}=\hbar^2/(2\mu R_\mathrm{vdw}^2)$ associated with the long range potential~\cite{Jones06}.  Table~\ref{table1} lists these for typical alkali species.  When $|E| \ll E_\mathrm{vdw}$, bound and scattering wave functions approach their long range asymptotic form for $R \gg R_\mathrm{vdw}$ and oscillate rapidly on a length scale small compared to $R_\mathrm{vdw}$ for $R \ll R_\mathrm{vdw}$.  When $E$ is much larger than $E_\mathrm{vdw}$, it is always a good approximation to make a semiclassical WKB connection between the short range and long range parts of the wave function, but when $E$ is small compared to $E_\mathrm{vdw}$,  a quantum connection is necessary, even for $s$-waves.  In such a case, $R_\mathrm{vdw}$ characterizes the distance range where the WKB connection fails~\cite{Julienne89}.   An alternative van der Waals length, $\bar{a}= 0.956 R_\mathrm{vdw}$, called the "mean scattering length" and which appears naturally in van der Waals theory, has been defined by Gribakin and Flambaum~\cite{Gribakin93}.  They also gave the the correction to universality in bound state binding energy due to the long range potential: $E_b=\hbar^2/(2 \mu (a-\bar{a})^2)$.

\begin{table}
\caption{ Characteristic van der Waals length and energy scales ($a_0=0.0529$ nm). }
\label{table1}
\begin{tabular}{ccccccl}
\hline\hline
Species &  $R_\mathrm{vdw}$ & $E_\mathrm{vdw}/k_B$ &
$E_\mathrm{vdw}/h$\\
     & ($a_0$) & (mK) & (MHz) \\
 \hline
 ${^6}$Li &  31.3 & 29.5  & 614 \\
 ${^{23}}$Na & 44.9  & 3.73   &   77.8 \\
 ${^{40}}$K &  64.9  & 1.03   & 21.4  \\
 ${^{87}}$Rb &  82.5  & 0.292  & 6.09\\
 ${^{133}}$Cs & 101  & 0.128  & 2.66\\
 \hline\hline
\end{tabular}\\
\end{table}

The key result from the MQDT analysis, derivable from formulas in the original papers~\cite{Mies84a,Mies84b}, is that the energy-dependent threshold width and shift can be written in the following factored form:
\begin{equation}
\frac{1}{2}\Gamma_n(E) =\frac{1}{2} \bar{\Gamma}_n C_\mathrm{bg}(E)^{-2}\hspace{2cm}
\delta E_n(E) =-\frac{1}{2}  \bar{\Gamma}_n \tan\lambda_\mathrm{bg}(E)\,,
\end{equation}
where $\bar{\Gamma}_n$ is independent of $E$ and $B$ and depends on the short range physics that determines the strength of the resonance.  The two functions $C_\mathrm{bg}(E)^{-2}$ and $\tan\lambda_\mathrm{bg}(E)$, as well as $\eta_\mathrm{bg}(E)$, are analytic functions of the entrance channel, or background, potential, and are completely given from the analytic van der Waals theory once $C_6$, reduced mass, and $a_\mathrm{bg}$ are specified.  When $E \gg E_\mathrm{vdw}$ such that the semiclassical WKB approximation applies at all $R$, the two MQDT functions take on the following limiting behavior~\cite{Mies84a,Mies84b,Julienne89}:
\begin{equation}
\lim_{E \gg E_\mathrm{vdw}} C_\mathrm{bg}(E)^{-2} = 1\hspace{2cm}
\lim_{E \gg E_\mathrm{vdw}}  \tan\lambda_\mathrm{bg}(E) = 0\,.
\end{equation}
On the other hand, when $E \ll E_\mathrm{vdw}$ so that the threshold law limiting behavior applies, then for the van der Waals potential~\cite{Mies00a}
\begin{equation}
\lim_{E\to 0} C_\mathrm{bg}(E)^{-2} = k\bar{a} \left ( 1+(1-r)^2\right ) \hspace{2cm}
\lim_{E \to 0}  \tan\lambda_\mathrm{bg}(E) = 1-r 
\end{equation}
\begin{equation}
 \frac{1}{2}\bar{\Gamma}_n = \frac{r}{1+(1-r)^2} \delta \mu \Delta_n
\end{equation}
where $r=a_\mathrm{bg}/\bar{a}$ represents the background scattering length in units of $\bar{a}$.  With these results the phase shift due to a Feshbach resonance takes on a remarkably simple form:
\begin{equation}
\eta(E,B)=\eta_\mathrm{bg}(E)-\tan^{-1}\left(\frac{\frac{1}{2} \bar{\Gamma}_n C_\mathrm{bg}(E)^{-2}}{E -\delta \mu(B-B_n)-\frac{1}{2}  \bar{\Gamma}_n \tan\lambda_\mathrm{bg}(E)}\right )
\label{eta.mqdt}
\end{equation}
The dependence on $B$ occurs only in the linear term in the denominator.  The dependence on the entrance channel is contained solely in the $\eta_\mathrm{bg}(E)$, $C_\mathrm{bg}(E)^{-2}$, and $ \tan\lambda_\mathrm{bg}(E)$ functions.  The strength of the resonant coupling is given by the single parameter  $\bar{\Gamma}_n$.  Numerical calculations show that phase shifts predicted by this formula are in superb agreement with full coupled channels calculations over wide ranges of $B$ and $E$ for typical resonances.

One pleasing aspect of MQDT theory is that $C(E)^{-1}$ has a simple physical interpretation~\cite{Mies84a,Mies84b}, described in the context of cold atom collisions by~\cite{Julienne89}.  The asymptotic entrance channel wave function for $R\gg\bar{a}$ is $f(R,E)=\sin(kR+\eta_\mathrm{bg}(E))/k^{1/2}$.  At short range, $R\ll\bar{a}$, it is convenient to write the wave function with a different normalization, that of the WKB semiclassical approximation, $\hat{f}(R,E)=\sin b(R,E)/k(R,E)^{1/2}$, where $k(R)=\sqrt{2\mu(E-V(R))/\hbar^2}$ is the local wavenumber.  Since the potential is quite large in magnitude at short range in comparison with $E\approx 0$, the $\hat{f}$ WKB function is essentially independent of energy and has a {\em shape} determined by the WKB phase $b(R)=\int k(R')dR'+\pi/4$, so that we can replace $\hat{f}(R,E)$ by $\hat{f}(R,0)$ when $R \ll \bar{a}$.  The WKB-assisted MQDT theory shows that the relation between $f(R,E)$ and $\hat{f}(R,E)$ (at all $R$) is $f(R,E)=C(E)^{-1}\hat{f}(R,E)$~\cite{Mies84a,Mies84b}.  Thus, when $R\ll\bar{a}$, the background channel wave function can be replaced by $C(E)^{-1}\hat{f}(R,0)$, so that, in particular, the coupling matrix element that determines the $E$-dependent width of the resonance can be written in factored form: $V_n(E)=\langle n|V|E\rangle=C(E)^{-1}\hat{V}_n$, where $\hat{V}_n$ depends only on the short range physics and is independent of $E$  and $B$ near threshold.  Thus, the short range physics, which depends on the large energy scale set by the deep short range potential, is separated from the long range physics and its very small energy scale near threshold.  Consequently, given $a_\mathrm{bg}$, the threshold properties depend only on the long range part of the potential.

\begin{figure}
  \includegraphics[height=.28\textheight]{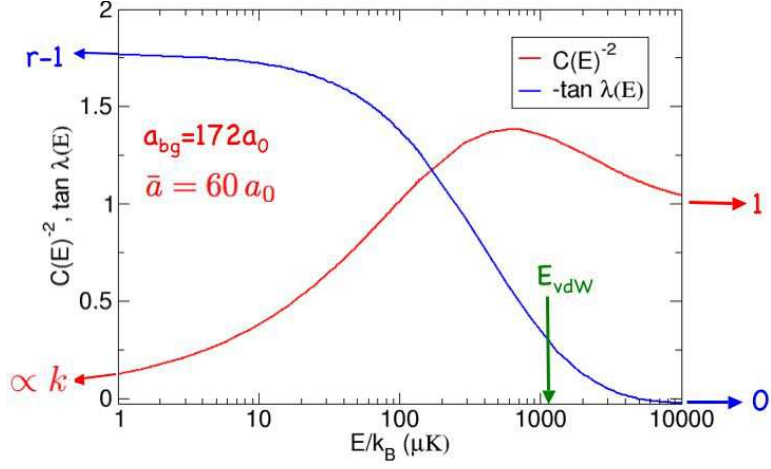}
  \caption{Calculated threshold background channel functions $C(E)^{-2}$ and $-\tan\lambda(E)$ for collisions of $^{40}$K atoms in the $a=\{F=9/2,M=-9/2\}$ and $b=\{F=9/2,M=-7/2\}$ states, showing the limiting behavior for energies that are large or small compared to the van der Waals energy $E_\mathrm{vdw}$ (arrow). }\label{fig1}
\end{figure}

\begin{figure}
  \includegraphics[height=.28\textheight]{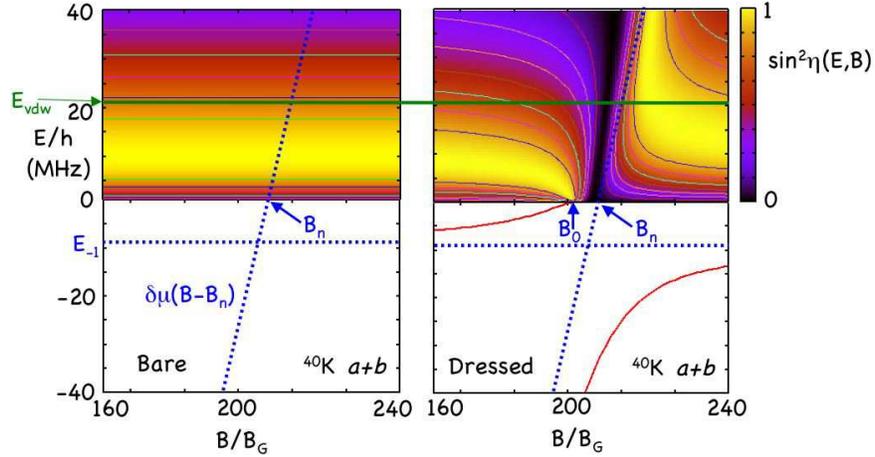}
  \caption{Scattering ($\sin^2\eta(E,B)$ for $E>0)$ and bound states (lines for $E<0$) near the $B/B_G=$202 $^{40}$K $a+b$ resonance, where $B_G=10^{-4}$T $=$ 1 Gauss, with the resonant coupling $V$ turned off (bare) or on (dressed).   Light-colored shading for $E>0$ implies scattering near the unitarity limit of the $S$-matrix.  The energy zero is chosen to be the energy of two motionless separated atoms at each $B$.  The energy scale of $E/h=\pm40$ MHz corresponds to a range of $E/k_B= \pm2$ mK, where $k_B$ is Boltzmann's constant.   The shift from $B_n$ to $B_0$, where  the dressed (solid line) bound state crosses threshold, is evident due to the avoided crossing between the last bound state of the background channel at $E_{-1}$ and the bare resonant state with $E_n=\delta\mu(B-B_n)$ and $\delta \mu/h = 23.5$ GHz$/$T.   The interference of the ramping bare resonance state with the background is evident for $E>0$.  A "sharp" resonance ($\Gamma_n(E) \ll E$) only emerges for $E \gg E_\mathrm{vdw}$~\cite{Nygaard06}. }\label{fig2}
\end{figure}

\begin{figure}
  \includegraphics[height=.30\textheight]{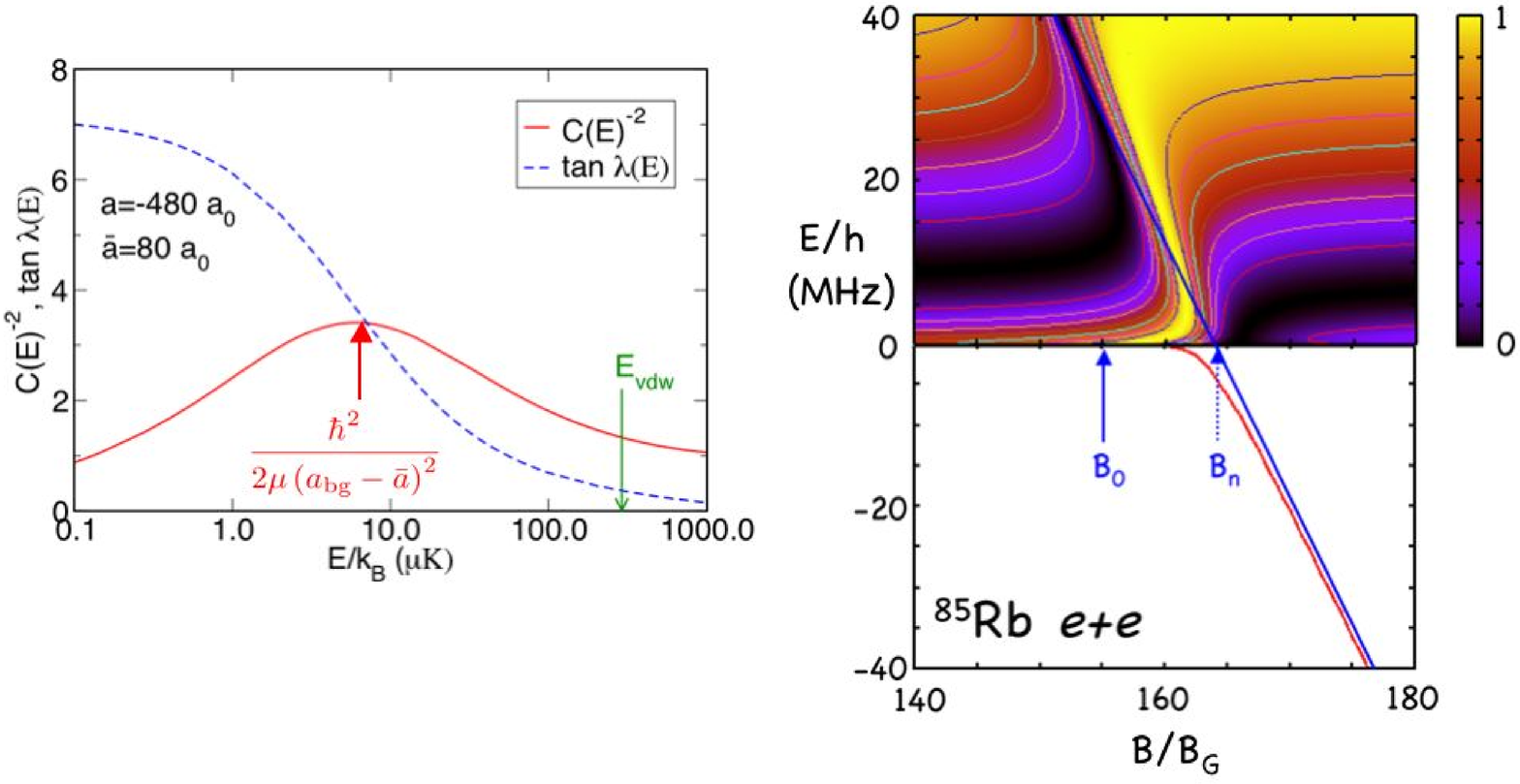}
  \caption{The left panel shows the $C(E)^{-2}$ and $\tan \lambda(E)$ functions for the background collision of two $^{85}$Rb atoms in the $e=\{F=2,M=-2\}$ state.  The van der Waals energy is $E_\mathrm{vdw}/k_B=0.3$ mK or $E_\mathrm{vdw}/h=6$ MHz. The right panel is analogous to Fig.~\ref{fig2}, with the same shading of $\sin^2\eta(E,B)$ for $E>0$ and with the bare and dressed bound states crossing threshold at $B_n$ and $B_0$ respectively ($B_G=10^{-4}$ T $= 1$ Gauss).  Marcelis {\it et al.}~\cite{Marcelis04} show a similar figure illustrating the large shift between $B_n$ and $B_0$. }\label{fig3}
\end{figure}

Figures ~\ref{fig1} and ~\ref{fig2} illustrate the threshold properties of a $^{40}$K resonance that has been used for experiments involving molecular Bose-Einstein condensation~\cite{Greiner03},  fermionic pairing~\cite{Regal04}, and reduced dimension~\cite{Moritz05}.  Figure ~\ref{fig3} shows similar results for a $^{85}$Rb resonance that has been used to make a stable condensate~\cite{Cornish00} and exhibit atom-molecule interconversion~\cite{Donley02}.  The $a_\mathrm{bg}$ is positive (negative) for $^{40}$K ($^{85}$Rb).   In both cases $|a_\mathrm{bg}| > \bar{a}$ and there is peak in $C(E)^{-2}$ near $E=\hbar^2/(2 \mu (a_{bg}-\bar{a})^2)$.  Recall that $C(E)^{-2}$ represents the squared amplitude enhancement of the short range wave function relative to the WKB wave function.  The peak value of $C(E)^{-2}$ is about 3.4 for the $^{85}$Rb case.  Note also that the shift $\delta E_n(E) \propto \tan \lambda(E)$  vanishes as $E$ increases above $E_\mathrm{vdw}$.   

The analytic van der Waals theory also gives the effective range of the potential~\cite{Gao00,Flambaum99}, and can be put into an angular-momentum independent form that predicts the bound states for different partial waves, given $a_\mathrm{bg}$, $C_6$, and reduced mass~\cite{Gao00,Gao01,Gao04a}.  For example, if $|a_{bg}|$ becomes very large compared to $\bar{a}$, there will be a $g$-wave bound or quasibound state near threshold.  This is the case for $^{133}$Cs and $^{85}$Rb, for example.  On the other hand, if $a_{bg}$ is close to $\bar{a}$ there will be a $d$-wave bound or quasibound state very close to threshold.  This is the case for $^{23}$Na, $^{87}$Rb, and $^{174}$Yb, all of which have $a$ slightly larger than $\bar{a}$ and all have quaisbound $d$-wave shape resonances close to threshold.

The van der Waals theory also permits a criterion for classifying resonances according to their open or closed channel character~\cite{Kohler06}.  Let us define a dimensionless resonance strength parameter $S=(a_\mathrm{bg}/\bar{a})(\delta \mu \Delta_n/E_\mathrm{vdw})$.  Let $Z$ be the norm of the closed channel component of the wave function of the near threshold dressed bound state ($1-Z$ is the norm of the open channel component).  Open channel dominated resonances have $S\gg 1$ and closed channel dominated ones have $S \ll 1$.  The bound states of the former are universal and have $Z \ll 1$ for a range of $|B-B_0|$ that is a significant fraction of $\Delta_n$; scattering states have $\Gamma_n(E) >E$ for $E < E_\mathrm{vdw}$, so that no resonance "feature" appears for $E <E_\mathrm{vdw}$.  The bound states for closed channel dominated resonances have only a small domain of universality near $B_0$ over a range that is small compared to $\Delta_n$, and have $Z\approx 1$ except over this narrow range; scattering states can have $\Gamma_n(E) \ll E$ for $E < E_\mathrm{vdw}$, so that sharp resonance features can appear in this range very close to threshold.  The $^{40}$K and $^{85}$Rb resonances described above, as well as the broad $^{6}$Li resonance near 834 G~\cite{Bartenstein05}, are open channel dominated.  The very narrow $^{6}$Li resonance near 543 G~\cite{Strecker03} and the $^{87}$Rb resonance near 1007 G~\cite{Thalhammer06,Volz06} are examples of closed channel dominated ones.  A good description of the two $^{6}$Li resonances has been given by ~\cite{Simonucci05}.

One application of the expression in Eq.~\eqref{eta.mqdt} is to provide an analytic form for the energy-dependent scattering length $a(E,B)=\tan^{-1}\eta(E,B)/k$.  This quantity is defined so as to give the scattering length in the limit $E\to0$ but can be used at finite $E$ to define an energy-dependent pseudopotential for finding interaction energies in strongly confined systems with reduced dimension~\cite{Blume02,Bolda02,Naidon06}.


Finally, it is worth noting that optical Feshbach resonances~\cite{Fedichev96,Bohn99} follow a similar theory to the magnetic ones.  An optical Feshbach is a photoassociation resonance~\cite{Jones06} where a laser with frequency $h\nu$ and intensity $I$ couples colliding ground state atoms to an excited bound state at frequency $h\nu_0$ relative to $E=0$ ground state atoms. Both the resonance width $\Gamma_n(E)$ and shift (contained in $h\nu_0$) are proportional to $I$.  The general expression for the resonant scattering length is
\begin{equation}
  a = a_\mathrm{bg} \left ( 1 - \frac{\Gamma_0}{E_0-i\frac{\gamma}{2}} \right ) = a_\mathrm{bg} -a_\mathrm{res} -ib_\mathrm{res} \,,
\end{equation}  
where, in the case of an optical resonance, $\Gamma_0=(l_\mathrm{opt}/a_\mathrm{bg})\gamma$, $E_0=-h(\nu-\nu_0)$, and $\gamma/\hbar$ is the spontaneous emission decay rate of the excited state.  This  compares with $\Gamma_0=\delta \mu \Delta_n$, $E_0=\delta \mu(B-B_0)$ and $\gamma/\hbar \approx 0$ for a typical magnetic resonance.  The optical resonance strength is characterized by the optical length $l_\mathrm{opt}$~\cite{Ciurylo05}, proportional to $I$ and the free-bound Franck-Condon factor $|\langle n|E\rangle|^2$.  The two-body decay rate coefficient due to decay of the resonance is $k_2=2(h/\mu)b_\mathrm{res}$.  Optical resonance are only practical when $|a_\mathrm{res}|\gg b_\mathrm{res}$, insuring that the decay rate is sufficiently small.  Assuming the detuning $\Delta=h(\nu-\nu_0)$ to be large in magnitude compared to $\gamma$, then $a_\mathrm{res} =l_\mathrm{opt} (\gamma/\Delta)$ and $b_\mathrm{res} =(l_\mathrm{opt}/2) (\gamma/\Delta)^2$.  Thus, the condition $\gamma/|\Delta| \ll 1$ ensures small decay, while the condition $l_\mathrm{opt} \gg |a_\mathrm{bg} -a_\mathrm{res}|$ is necessary in order to make a meaningful change in $a_\mathrm{bg}$.

Satisfying the above conditions is difficult for strongly allowed optical transitions for which $\gamma$ is large, since very large detuning from atomic resonance is necessary in order to suppress losses, and then $l_\mathrm{opt}$ is too small for reasonable intersity $I$; alternatively changing $a_\mathrm{bg}$ is accompanied by large losses.  These problems can be eliminated by working with narrow intercombination line transitions of species such as Ca, Sr, or Yb~\cite{Ciurylo05,Zelevinsky06,Tojo06}.  Quite large values of $l_\mathrm{opt}$ can be achieved for levels very close to the atomic dissociation limit of the molecular excited state, and it appears to be feasible to use such resonances to control scattering lengths~\cite{Zelevinsky06}.



%





\end{document}